\documentclass{aa}
\usepackage{graphicx}
\usepackage{subcaption}
\usepackage[varg]{txfonts}
\usepackage{hyperref}
\usepackage{natbib}
\bibpunct{(}{)}{;}{a}{}{,}
\usepackage{wasysym}
\usepackage{siunitx}
\usepackage{longfigure}
\usepackage{color}
\usepackage{acronym}
\usepackage{float}
\usepackage{mathrsfs}
\usepackage[cal=boondox]{mathalfa}
\usepackage{dutchcal}
\usepackage{enumitem}
\usepackage[normalem]{ulem}

\begin{document} 

   \title{HE2159-0551: a very metal-poor peculiar low Ba giant -- a candidate from the LMS-1/Wukong}
   \titlerunning{HE2159-0551: a VMP peculiar low Ba giant -- candidate for LMS-1/Wukong}
   \author{A. Schichtel
         \and
         D. H. Plonka
          \and
          C. J. Hansen
          }
   \institute{Goethe-Universität, Institute for Applied Physics, Max-von-Laue-Straße 1, 60438 Frankfurt am Main, Germany\\
              \email{schichtel@em.uni-frankfurt.de}
             }

   \date{Received DAY MONTH 2025 / Accepted DAY MONTH 2025}

  \abstract
   {}
   {We present a comprehensive spectroscopic and kinematic analysis of the very metal-poor ([Fe/H] = $-$2.60 $\pm$ 0.20 dex) giant star \object{HE2159-0551}. By investigating the star's chemodynamic characteristics, we seek to address
   its formation, evolution, and role in the chemical enrichment of the early cosmos and the possible connection to known merger events.}
   {From high-resolution data we perform a one dimension, local thermodynamic equilibrium analysis using PyMoogi. We conduct a detailed abundance analysis of 23 elements (C, N, O, Na, Mg, Al, Si, K, Ca, Sc, Ti, V, Cr, Mn, Fe, Co, Ni, Zn, Sr, Y, Zr, Ba, and Eu), allowing us to derive abundances or place limits. Finally, we compute orbital parameters to investigate the kinematics and thus the nucleosynthetic origin of \object{HE2159-0551}.}
   {The analysis yields significant insights into the chemical composition of \object{HE2159-0551}, highlighting its peculiarities among very metal-poor stars. The star shows signs of internal mixing and a peculiar abundance pattern, particular regarding the heavy elements. We find a low Ba abundance (s-process element) and a suppressed contamination of r-process elements, even though r-process enrichment may be expected in very metal-poor stars. The orbital parameters and kinematic properties indicate that \object{HE2159-0551} is a thick disc star, connected to the old and metal-poor LMS-1/Wukong progenitor resulting from an early merger.}
   {Our kinematic analysis suggests a potential connection of \object{HE2159-0551} to the merger LMS-1/Wukong. We conclude that the nucleosynthesis processes responsible for the star’s enrichment in heavy elements are different from those observed in many other metal-poor stars as neither the main r- nor s-process can explain the derived abundance pattern. A possible weak-r or $\nu$p-process forming Sr-Zr but not Ba might have mixed into a low level of underlying r-process material in the low-mass LMS-1/Wukong system.}


   \maketitle
%

\section{Introduction}\label{section:introduction}
Very metal-poor (VMP) stars, which exhibit a peculiar abundance pattern, are crucial for advancing our understanding of stellar and Galactic chemical evolution (GCE) and early nucleosynthesis. In particular, the rapid neutron-capture process (r-process) is responsible for producing half the heavy elements beyond iron early on. Its study is essential for unraveling the complexities of element formation in the cosmos \citep{burbidge1957synthesis,christlieb2004hamburg,arcones2023origin}. 
Stellar abundances of neutron-capture (n-capture) elements are particularly important for identifying the astrophysical sites of nucleosynthesis. The r-process forms Eu and is often linked to the explosive conditions of rare core-collapse (CC) supernovae (SNe) or neutron-star-mergers (NSMs -- or compact mergers) \citep[e.g.,][]{barklem2005hamburg,winteler2012magnetorotationally,hansen2014many,Hotokezaka_2018,hansen2018r,Cowan_2021}. The detection of the r-process signature in the kilonova associated with the NSM event GW1708178 provides compelling evidence for a NSM to be a site of the r-process nucleosynthesis; based on one single event to date \citep[e.g.,][]{2017Abbott,2019watson}. The weak r-process is a crucial nucleosynthesis process that occurs in environments such as neutrino-driven winds from CC SNe. This process might be responsible for the formation of certain elements up to Z $\sim$ 50 and plays a significant role in the chemical evolution \citep{montes2007nucleosynthesis,hansen2012silver,arcones2014nucleosynthesis}. The $\nu$p-process occurs in the neutrino wind from a hot proto-neutron star following a SN explosion, providing a crucial environment for the synthesis of elements beyond Fe \citep{Frohlich2006,frohlich2012reaction}. As the proto-neutron star cools and emits neutrinos, these particles interact with the surrounding stellar matter, creating proton-rich conditions that facilitate nucleosynthesis typically producing elements around Fe and beyond up to Sr-Zr, possibly reaching Ru \citep{Frohlich2006}. 
Understanding the weak r- and $\nu$p-process is essential for elucidating the origins of the elements we observe and detailed abundance patterns in metal-poor (MP) stars help us separate and understand these processes in more detail.

Following the classification from \citet{beers2005discovery}, VMP stars are defined as having [Fe/H] < $-2.00$ dex. The abundance patterns of these stars can vary significantly (see e.g., the r-poor and r-rich stars analyzed by \citet{honda2006neutron} and \citet{sneden2003extremely}, respectively), offering valuable insights into the stars' origin and evolutionary histories \citep{franccois2007first,hansen2012silver}. These variations in r-process elements challenge existing theories by illuminating the history of chemical enrichment and mixing processes throughout a star's life cycle. As stars ascend the giant branch, they can experience significant internal mixing processes and nucleosynthesis, which can alter their surface compositions \citep[e.g.,][]{korn2006probable,korn2008he,pedersen2021internal,FernandesdeMelo2024}. Despite the potential insights from studying these stars, our understanding of the specific evolutionary pathways leading to unusual abundance patterns remains limited.
Stars with enhanced C, N, and Ba abundances \citep[e.g.,][]{Hansen_2016,Hansen_2019,Zhang_2023}, the latter typically formed by the slow neutron-capture process (s-process), often reflect the products and mass-transfer of asymptotic giant branch (AGB) stars. 
However, some VMP giants display peculiar abundance patterns that deviate from established trends, raising critical questions about the mechanisms behind these anomalies \citep[e.g.,][]{sneden2003extremely,honda2006neutron}. 

Significant fractions of the Milky Way’s stellar halo are composed of stars that originated in low-mass companions, ranging from dwarf spheroidals and ultra-faint dwarf galaxies to bound systems like globular clusters and were later accreted through gravitational interactions and mergers \citep[e.g.,][]{Koppelmann_2019b,Aguado_2021a,Matsuno_2022b,Matsuno_2022a}. These accretion events are key to unraveling our Galaxy’s early assembly history, since each merger leaves behind distinctive chemical and dynamical signatures in the halo’s stellar populations. Recent studies using \textit{Gaia} data and large-scale spectroscopic surveys, e.g., APOGEE \citep{2017AJ....154...28B} or GALAH \citep{buder2024galahsurveydatarelease}, have revealed numerous stellar substructures associated with distinct past accretion events \citep[e.g.,][]{Helmi_2018,Naidu_2020,Yuan_2020}. Among these newly identified substructures, the low-mass-stream-1 (LMS-1), a possible low-mass extension of Wukong\footnote{In this work, we consider LMS-1 and Wukong to be components of the same underlying accretion structure. This assumption is supported by their similar orbital properties and chemical patterns, which we explore in Sect. \ref{section:discussion-conclusion}. We therefore refer to it as LMS-1/Wukong.}, stands out as a particularly interesting component due to its exceptionally low metallicity and polar orbital configuration \citep{Yuan_2020,Naidu_2020,Malhan_2021}.

In this paper, we focus on the analysis of the chemically peculiar, VMP giant star \object{HE2159-0551}, whose kinematic and chemical properties suggest a connection to the LMS-1/Wukong merger event and we find it exhibits an abundance pattern that deviates from the expected heavy element ($Z>30$) trends observed in similar stars. Through a detailed chemodynamical analysis, we explore its possible association with LMS-1/Wukong, thereby refining our understanding of the interplay between accretion processes, Galactic dynamical evolution, and early nucleosynthesis. Understanding the origin of abundance peculiarities is crucial for refining models of stellar evolution and chemical enrichment in the early galaxy.

In Sect. \ref{section:observations}, we present a brief overview of the observational data. Section \ref{section:radial-velocity} details the radial velocity (RV) computations, followed by the stellar parameters in Sect. \ref{section:stellar-parameters}. Section \ref{section:kinematics} describes the star's kinematics, while Sect. \ref{section:abundances} focuses on the chemical abundances of 23 elements. Section \ref{section:discussion-conclusion} presents the results and concludes the study.

\section{Observations}\label{section:observations}
We use high-resolution spectra obtained with UVES (Ultraviolet and Visual Echelle Spectrograph) at the VLT (Very Large Telescope), observed over five non-consecutive nights. The signal-to-noise ratio (SNR) is approximately 200 around 6060 \AA{} and 30 around 3440 \AA; for details on the spectra see Table \ref{tab:rvs} in the appendix. The spectra cover a wavelength range from approximately 3000 \AA{} to $\sim$ 9500 \AA{} with some gaps due to different arms and dichroics. The central wavelengths are 346 nm and 390 nm for the blue arm and 564 nm, 580 nm, and 760 nm for the red arm. The UVES spectra undergo automatic reduction using the UVES pipeline, which includes bias correction, flat-fielding, sky subtraction, order merging, wavelength calibration, and background subtraction \citep{dekker2000design}. The star has a mean G magnitude of 12.0582 $\pm$ 0.0002 \citep{2023GaiaDR3} and the coordinates (J2000) are RA: 22 02 16.3602355800, Dec: $-$05 36 48.534816816.

\section{Radial velocity}\label{section:radial-velocity}
We employ IRAF\footnote{\url{https://iraf-community.github.io/}} \citep{tody1986iraf,tody1993iraf,2024arXiv240101982F} to compute the RV by analyzing each individual spectrum before coadding. We identify the Balmer lines in IRAF (using \texttt{rvidlines}), compare the observed wavelengths to their rest frame wavelengths to determine the RV, and later correct for the barycentric velocity. We shift the files based on this value. Table \ref{tab:rvs} presents the RV value for each individual file and we find a final RV value of $-$106.40 $\pm$ 1.20 km\,s$^{-1}$ as an average for all spectra. We compare this computation to a second method by using a cross-correlation method against the spectrum from July 28, 2015 (see Table \ref{tab:rvs}), which results in a RV of $-$107.00 $\pm$ 4.00 km\,s$^{-1}$, agreeing with the RV result of our initial analysis\footnote{The uncertainty of the radial velocities (RVs) can be computed using the quadratic sum, as the values are independent of one another. The standard deviation of the RV is 1.20 km\,s$^{-1}$ for the individual, not combined spectra. The estimated standard error of the mean is 0.40 km\,s$^{-1}$.}. Table \ref{tab:rvs} reveals small differences in the individual RV values, which may be attributed to a slight reduction discrepancy. Notably, the largest discrepancy in RVs is observed in one file observed on July 28, 2015. The cross-correlation method yields $-$117.36 km\,s$^{-1}$, while the single-file analysis results in $-$104.84 km\,s$^{-1}$ after applying the barycentric correction. There are sufficient lines available for cross-correlation and the influence of telluric lines on the results is expected to be minimal as the spectra undergo sky subtraction, see Sect. \ref{section:observations}. 

The \citet{2016Gaia,2023GaiaDR3} report a RV of $-$106.38 $\pm$ 0.80 km\,s$^{-1}$, with the uncertainty calculated as the standard error of the mean. \citet{hansen2018r} give a value of $-$106.65 $\pm$ 0.88 km\,s$^{-1}$, while \citet{hansen2015elemental} provide a RV value of $-$131.30 $\pm$ 0.80 km\,s$^{-1}$. This RV value is comparable to the values we compute prior to applying the barycentric correction of 25.21 km\,s$^{-1}$, see Table \ref{tab:rvs}. This indicates that the RV in \citet{hansen2015elemental} is given without the barycentric correction. Overall, our computed RV aligns well with the literature, see Table \ref{tab:rvs-lit}. Based on the above we exclude binarity.

\begin{table}
	\centering
	\caption{RV literature values and the uncertainties based on \citet{hansen2015elemental,hansen2018r} and the \citet{2016Gaia,2023GaiaDR3} compared to this work (single-file analysis (shifting) and cross-correlation (CC) method).}
	\label{tab:rvs-lit}
	\begin{tabular}{ll}
	\hline\hline
        Data origin & RV [km\,s$^{-1}$] \\
        \hline
		This work\textsubscript{final, shifting} & $-$106.40 $\pm$ 1.20\\
        This work\textsubscript{final, CC} & $-$107.00 $\pm$ 4.00\\
        \citet{hansen2015elemental} & $-$131.30 $\pm$ 0.80 \\ 
        \citet{hansen2018r} & $-$106.65 $\pm$ 0.88 \\ 
       \citet{2016Gaia,2023GaiaDR3} & $-$106.38 $\pm$ 0.80 \\
		\hline
	\end{tabular}
\end{table}

\section{Stellar parameters}\label{section:stellar-parameters}
We determine the stellar parameters spectroscopically using 20 Fe I and 14 Fe II lines from a known line list across a broad wavelength range. We list the individual lines and references in Table \ref{tab:abundances}. For blended lines, we employ the deblending function in IRAF, as neglecting this step can result in significantly inaccurate abundance measurements. We generate our atmospheric model by interpolating Kurucz ATLAS9 models \citep{kurucz1970atlas,castelli2004new}, which we then use as input for PyMoogi\footnote{\url{https://github.com/madamow/pymoogi}, based on the 2019 version of MOOG.}. The results of the stellar parameters are presented in Table \ref{tab:stellarparametrs-lit}, alongside the literature data from \citet{hansen2015elemental}. We determine the statistical uncertainty on the metallicity as the standard deviation stemming from the Fe I lines using the final adopted atmospheric model in PyMoogi. For the other stellar parameters (microturbulence, surface gravity, and temperature) we compute detailed uncertainties (see Sect. \ref{sec:StelllarParametersUncertanties} in the appendix).

\begin{table}
	\centering
    \caption{This work's stellar parameters compared to \citet{hansen2015elemental}.}
	\label{tab:stellarparametrs-lit}
	\begin{tabular}{llrr}
	\hline \hline
        Parameter & Unit & This work & \citet{hansen2015elemental} \\
        \hline
        T\textsubscript{eff} &  [K] & 5000 $\pm$ 140 & 4800 $\pm$ 100 \\
        log g & [dex] & 1.80 $\pm$ 0.40 & 1.50 $\pm$ 0.30 \\
        $\xi$ & [km\,s$^{-1}$] & 2.40 $\pm$ 0.30 & 2.10 $\pm$ 0.30 \\
        $[$Fe/H$]$ &  [dex] & $-$2.60 $\pm$ 0.20 & $-$2.80 $\pm$ 0.20 \\
		\hline
	\end{tabular}
\end{table}

Following the methods outlined by \citet{mucciarelli2013gala} and \citet{hansen2011diss}, we derive the effective temperature by using Fe I and Fe II lines. We aim for a flat slope in the abundance measurements of the examined lines (for both neutral and ionized Fe lines), regardless of the excitation potential of the lines. We compute an effective temperature of 5000 $\pm$ 140 K.

We use the abundance of Fe I to determine the star's metallicity, based on the equivalent widths (EWs). While metallicity can be assessed using Fe I lines alone, these lines are more susceptible to non-local thermodynamic equilibrium (NLTE) effects compared to Fe II lines \citep{bergemann2012non}. However, the number of detectable Fe II lines is limited, which is why the determination of the metallicity by Fe II lines alone could be statistically falsified. We derive [Fe/H] = $-$2.60 $\pm$ 0.20 dex in one dimension, local thermodynamic equilibrium. The uncertainty reflects the systematic uncertainty of the Fe abundance. We double-check the metallicity by synthesizing the Fe lines listed in Table \ref{tab:abundances}.

We require ionization equilibrium between the Fe I and Fe II lines to determine the surface gravity. However, the surface gravity estimate may be biased, as Fe I is a non-dominant species in cool giants and could be affected by NLTE conditions, potentially introducing systematic errors in these measurements \citep{hansen2011diss,bergemann2011nlte}. We compute a log g of 1.80 $\pm$ 0.40 dex. 

We assume that Fe I lines should provide consistent Fe abundance measurements regardless of their individual strengths, implying no dependency between line strength and Fe abundance \citep{hansen2011diss,mucciarelli2013gala}. We compute a microturbulence of 2.40 $\pm$ 0.30 km\,s$^{-1}$. Following the empirical formula given by \citet{mashonkina2017formation} we derive a microturbulence of 1.95 km\,s$^{-1}$. \citet{mashonkina2017formation} conclude that the empirical formula should only be used for extremely and ultra MP stars, while \object{HE2159-0551} is a VMP star. Therefore, we keep the 2.40 km\,s$^{-1}$ as the microturbulence.

We include the literature values of the stellar parameters from \citet{hansen2015elemental} in Table \ref{tab:stellarparametrs-lit}, alongside the values we derive. These values fall within the combined uncertainties. Notable differences, particularly in the effective temperature, likely arise from the different methods employed to derive the stellar parameters. We utilized a spectroscopic approach, while \citet{hansen2015elemental} employed a fitting method that matches spectrophotometric observations with model atmospheres. Following \citet{mucciarelli2020facing}, the differences between the stellar parameters of giant stars derived from spectroscopy and photometry increase with decreasing metallicity. The spectroscopic temperature for a star with [Fe/H] = $-$2.60 dex could be lower by around 350 K \citep{mucciarelli2020facing}. We determine a spectroscopic effective temperature that is 200 K higher than the photometric value reported by \citet{hansen2015elemental}. 

Automated tools like xiru\footnote{\url{https://github.com/arthur-puls/xiru}} \citep{AlencastroPuls_phd}, a Python wrapper for the MOOG code, or ATHOS\footnote{\url{https://github.com/mihanke/athos}} (A Tool for HOmogenizing Stellar parameters, \citet{hanke2018athos}) have the potential to enhance precision; however, these tools often have limitations when it comes to individual parameters. We employed xiru to derive the stellar parameters based on the EW measurements. This analysis yielded the following results: T\textsubscript{eff} = 4785.40 $\pm$ 204.80 K, $\log$ g = 0.72 $\pm$ 1.11 dex, [Fe/H] = $-$2.95 $\pm$ 0.17 dex, and v\textsubscript{t} = 3.02 $\pm$ 0.36 km\,s$^{-1}$. Notably, all derived stellar parameters are within combined uncertainties when comparing to our results presented in Sect. \ref{section:stellar-parameters}, reinforcing the reliability of our measurements. ATHOS returns similar or lower values than xiru, still within combined uncertainties compared to our results.

\section{Kinematics}\label{section:kinematics}

\subsection{Computing orbital parameters}\label{subsection:computing_orbital_parameters}
To compute the orbital parameters of an object, e.g., actions and energies $(\mathbf{J},E)$, we require the data of the complete 6D phase-space measurements of our object. For the 2D sky position $(\alpha, \delta)$ and the 2D proper motion $(\mu^*_\alpha \equiv \mu_\alpha \, cos(\delta), \, \mu_\delta)$ along with the associated uncertainties, we use the data for \object{HE2159-0551} from the \textit{Gaia} DR3 data set \citep{2016Gaia,2023GaiaDR3}, while the parameters for the heliocentric distance $(D_\odot)$ are based on the \citet{BailerJones+} catalog. The computed value for the RV is discussed in Sect. \ref{section:radial-velocity}. The used 6D phase-space values and their uncertainties are shown in Table \ref{tab:6d space-phase values} in the appendix.

We adopt the Galactic potential of \citet{McMillan2017} to compute the orbits and orbital parameters of \object{HE2159-0551}. This time-independent, static, and axissymmetric model of the Milky Way contains a bulge, disk components, and the NFW (Navarro-Frenk-White) halo. To set the \citet{McMillan2017} potential and to integrate the orbits, we make use of the \texttt{galpy}\footnote{\url{http://github.com/jobovy/galpy}} module \citep{Bovy_2015}. For every calculated orbit in this paper, we use the same total integration time of 10 Gyr, with time steps of 1 Myr. For further details we refer to Plonka et al. (in prep.). We take solar and local standard of rest (LSR) data from \citet{BennetBovy2018} and \citet{Schoenrich2010}, respectively.

Starting from GALAH\,DR4 \citep{buder2024galahsurveydatarelease}, we select a comparison sample of 480\,791 stars that satisfy the recommended quality flags (\texttt{flag\_sp = 0}, \texttt{flag\_sp\_fit = 0}, \texttt{flag\_red = 0}, \texttt{flag\_fe\_h = 0}), ensuring reliable parameter and abundance estimates. These stars span effective temperatures in the approximate range 4000–7500\,K with a median near 5550\,K, and cover a metallicity interval of  $-3.00$ dex $\lesssim \mathrm{[Fe/H]} \lesssim +0.75$ dex, though the MP tail below $-$1.50 dex is relatively sparsely sampled. The sample includes both dwarf and giant stars. For all selected stars, we compute orbital parameters in the same way as for \object{HE2159‑0551}, enabling a direct chemodynamical comparison.

In order to quantify the uncertainties in the derived orbital parameters, we employ a Monte Carlo sampling technique. First, the measured quantities, namely the RV, distance, and proper motions, along with their respective uncertainties, are incorporated into a covariance matrix. The diagonal elements of this matrix represent the variances of the individual parameters ($\sigma^2_{v_{\rm rad}}$, $\sigma^2_{\rm dist}$, $\sigma^2_{\rm pmRA}$, $\sigma^2_{\rm pmDec}$), while the off-diagonal elements account for correlations between pmRA and pmDec. We then sample from a multivariate normal distribution with the vector of measured values as the mean and the constructed covariance matrix as the scatter. By drawing a large number of samples ($n=10^5$ for \object{HE2159-0551} and $n=1$ for background stars), we generate variations of the input data that reflect the measurement uncertainties.

For each sampled set of input values, an orbit is integrated using our orbit integration code, yielding a distribution of orbital parameters such as energy, eccentricity, and actions. To estimate the associated uncertainties, we use the 16th and 84th percentiles of the resulting distributions, corresponding to a 68 \% confidence interval around the median. This approach provides a robust measure of parameter uncertainties that accounts for the asymmetry in their distributions, rather than relying solely on standard deviations. We show the calculated values with their associated uncertainties for \object{HE2159-0551} in Table \ref{tab:orbital_parameters}.

\begin{table}
    \centering
    \renewcommand{\arraystretch}{1.5}
    \caption{Calculated orbital parameters for \object{HE2159-0551}.}
    \begin{tabular}{lll}
    \hline \hline
            Parameter & Unit & \object{HE2159-0551}\\
        \hline
        E & [$\text{km}^2$\,s$^{-2}$] & $-162\,906.15^{+21\,919.05}_{-12\,653.19}$\\
        $\text{J}_{\text{r}}$ & [kpc km\,s$^{-1}$] & $151.48^{+242.32}_{-109.29}$\\
        $\text{J}_\phi$ & [kpc km\,s$^{-1}$] & $855.08^{+859.82}_{-801.34}$\\
        $\text{J}_{\text{z}}$ & [kpc km\,s$^{-1}$] & $318.94^{+636.27}_{-226.91}$\\
        $\text{L}_{\text{z}}$ & [kpc km\,s$^{-1}$] & $855.08^{+859.82}_{-801.34}$\\
        $\text{R}_{\text{Apo}}$ & [kpc] & $8.96^{+5.46}_{-1.15}$\\
        $\text{R}_{\text{Peri}}$ & [kpc] & $4.77^{+2.80}_{-2.85}$\\
        $\mathcal{e}$ & & $0.39^{+0.28}_{-0.20}$\\
        $\text{Z}_{max}$ & [kpc] & $5.80^{+4.53}_{-3.33}$\\
        $\text{P}_r$ & [Gyr] & $0.13^{+0.08}_{-0.03}$\\
        $\text{U}_{LSR}$ & [km\,s$^{-1}$] & $-16.85^{+69.34}_{-62.55}$\\
        $\text{V}_{LSR}$ & [km\,s$^{-1}$] & $-99.59^{+99.99}_{-131.22}$\\
        $\text{W}_{LSR}$ & [km\,s$^{-1}$] & $44.65^{+89.63}_{-94.94}$\\
    \hline    
    \end{tabular}
    \label{tab:orbital_parameters}
\end{table}

\subsection{HE2159-0551 -- currently a thick disc star}
\begin{figure}
    \centering
    \includegraphics[width=\linewidth]{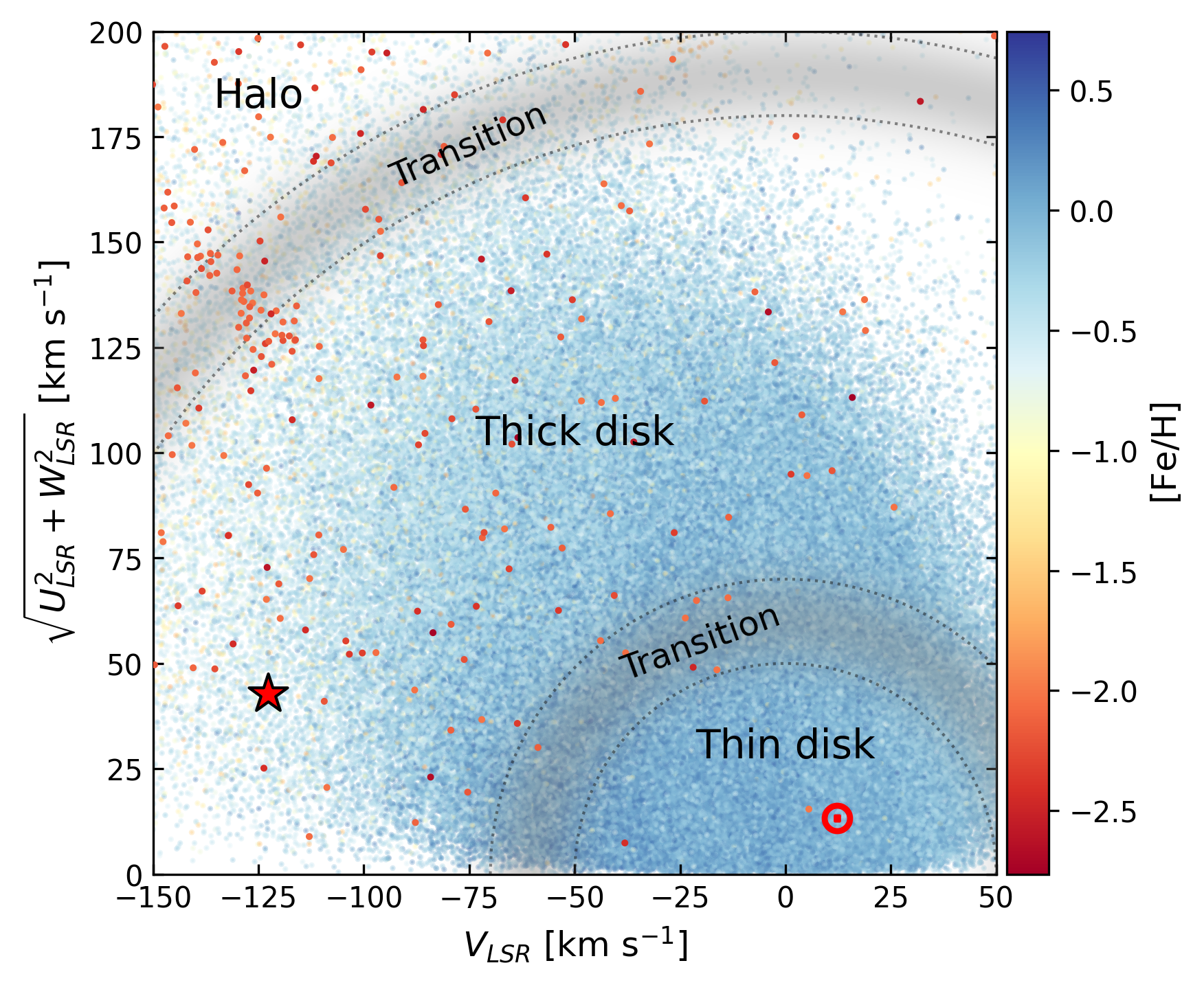}
    \caption{Toomre diagram showing \object{HE2159-0551} (red star); Sun is indicated as red circle. The dotted lines represent contours of constant total space velocity $v_{\mathrm{tot}}$. Stars from the GALAH background sample \citep{buder2024galahsurveydatarelease} with [Fe/H] $< - 2.00$ dex are highlighted.}
    \label{fig:toomre_diagram}
\end{figure}

Following \citet{Bensby2003,Bensby2014}, we divide the Milky Way into three Galactic components: stars with thick disc-like orbits \citep[$70 \lesssim v_{tot} \lesssim 180$ km\,s$^{-1}$;][]{Nissen2004}, stars with thin disc-like orbits ($v_{tot} \equiv (U^2_{LSR} + V^2_{LSR} + U^2_{LSR})^{1/2}$ $\lesssim$ $50$ km\,s$^{-1}$), and stars with halo-like orbits ($v_{tot} > 200$ km\,s$^{-1}$). This method uses a probabilistic separation and assumes that the Galactic velocities $(U_{LSR}, V_{LSR}, W_{LSR})$ of these components follow different Gaussian distributions. We show the resulting Toomre diagram of \object{HE2159-0551} in Fig. \ref{fig:toomre_diagram}, and in combination with its orbital properties, \object{HE2159-0551} currently seems to belong to the thick disc population of the Milky Way.

\subsection{Investigating the origin of HE2159-0551}\label{subsec:kinematics-origins}

\begin{figure*}
    \centering
    \includegraphics[width=\linewidth]{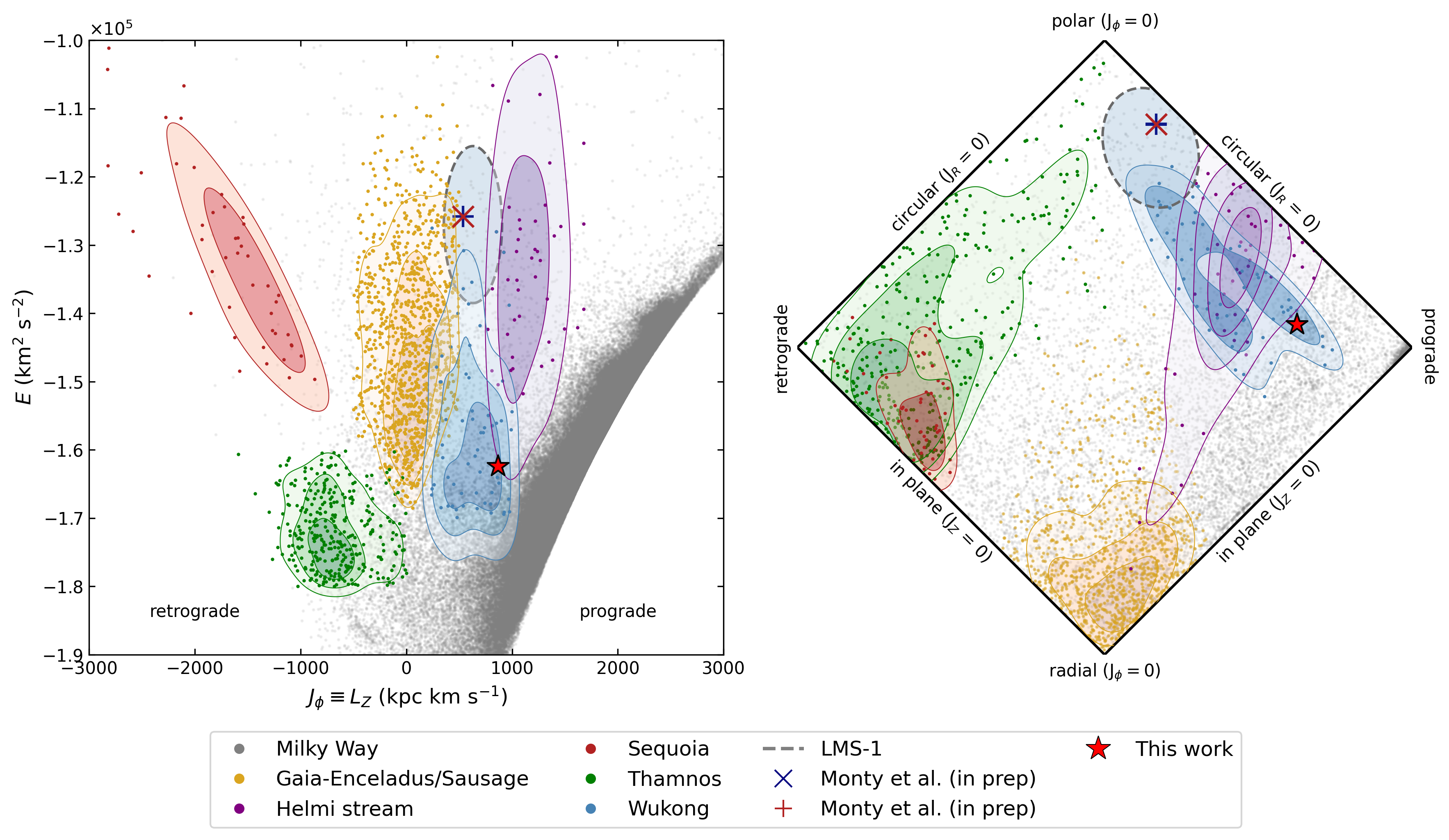}
    \caption{Orbital properties of \object{HE2159-0551} (red star) compared to known stellar substructures in the Milky Way. 
        Left: $E$–$L_z$ diagram showing orbital energy vs. angular momentum. Colored points and contours (based on a Gaussian density distribution) indicate the location of accreted populations from the literature. Specifically, Gaia-Enceladus/Sausage \citep{Belokurov_2018, Helmi_2018}, Sequoia \citep{Myeong_2019}, Thamnos \citep{Koppelman_2019_Thamnos}, Helmi stream \citep{Helmi_1999}, LMS-1 \citep{Yuan_2020}, and Wukong \citep{Naidu_2020}. \object{HE2159-0551} lies within the Wukong selection region, but at the prograde edge of the structure.
        Right: Action diamond diagram with $x = J_\phi / J_\mathrm{tot}$ and $y = (J_z - J_R) / J_\mathrm{tot}$, where $J_\mathrm{tot} = J_R + |J_\phi| + J_z$. \object{HE2159-0551} again appears within the Wukong region.}
    \label{fig:Lz-E_action_diamond}
\end{figure*}

To further examine the origin of \object{HE2159-0551}, we analyze its position in the E-L$_z$ plane. Because orbital energy and the $z$-component of angular momentum are integrals of motion, stars originating from the same accretion event naturally share similar values of these quantities, even when scattered throughout the Galaxy \citep[e.g.,][]{Helmi_2000,Gomez_2010,Simpson_2019}. As a first approximation, we use the identification of the substructures in the stellar halo, as stated in \citet{Naidu_2020} and \citet{Horta2022}. Using the previously calculated orbital parameters for \object{HE2159-0551} and the background stars (see Sect. \ref{subsection:computing_orbital_parameters}), we plot the E-L$_z$ plane in Fig. \ref{fig:Lz-E_action_diamond} (left). We note that \object{HE2159-0551} fits in the following selection criteria for the LMS-1/Wukong\footnote{Note that \citet{Horta2022} refers to the Wukong merger, originally identified by \citet{Naidu_2020}, using the name LMS-1.} merger from \citet{Horta2022}: $0.20 < L_z < 1$ [$10^3$ kpc kms$^{-1}$], $-1.70 < E < -1.20$ [$10^5 \, \text{km}^{2}\text{s}^{-2}$], [Fe/H] $< -1.45$ dex, $0.40 < \mathcal{e} < 0.70$, and $|Z| > 3$ [kpc]. Adopting the selection criteria from \citet{Horta2022}, we identify 62 LMS-1/Wukong stars in our background sample.

LMS-1 is a stellar substructure that was recently discovered by \citet{Yuan_2020}. It comprises MP stars that manifest themselves as an overdensity close to the lower edge of the Gaia-Enceladus/Sausage (GES) merger in the E-L$_z$ plane. This population is subsequently analyzed in detail by \citet{Naidu_2020}, who name it Wukong and highlight its slightly prograde orbital motion in a polar orientation. \citet{Malhan_2022} conclude that it represents the most MP accretion event identified so far in the Milky Way, characterized by a metallicity distribution function (MDF) extending down to [Fe/H] $\approx -3.40$ dex.  \citet{Limberg_2024} further refine the dynamical characterization of LMS-1/Wukong. Combining kinematics and abundances we find a chemodynamical selection important and stick to the definition from \citet{Horta2022}. Based on the Mg and Fe abundances and the overlap in the E-L$_z$ plane of LMS-1/Wukong and the Helmi-Stream, it is possible that these two substructures could be linked, where LMS-1/Wukong constitutes the more MP component of the Helmi-Stream. However, \cite{Naidu_2020} suggest it to be an independent substructure based on the metallicity peak in the MDF of the stars including their definition in the E-L$_z$ plane. In addition to \cite{Naidu_2020} and \citet{Horta2022}, we also show an alternative selection for LMS-1/Wukong, based on unpublished criteria from Monty et al. (in prep.), indicated by the dashed ellipse in Fig. \ref{fig:Lz-E_action_diamond}. This region encloses the typical locus of their LMS-1/Wukong candidates in the E-L$_z$ space and is manually defined to capture their dynamical clustering. 

Several extremely MP stellar streams have since been linked to LMS-1/Wukong. In particular, the streams C19 ([Fe/H] $=-3.38 \pm 0.06$ dex; \citealt{Martin_2022a}), Sylgr ([Fe/H] $=-2.92\pm0.06$ dex; \citealt{Roederer_Gnedin_2019}), and Phoenix ([Fe/H] $=-2.70\pm0.06$ dex; \citealt{Wan_2020}) are identified as belonging to this merger. Additionally, the stellar streams Indus and Jhelum have preliminary been proposed as being associated with LMS-1/Wukong by \cite{Bonaca_2021}. While the association of Indus with LMS-1/Wukong is further supported by \cite{Malhan_2021}, the same study cast doubt on Jhelum’s membership. Considering that both Indus and Jhelum are debris of disrupted dwarf galaxies \citep{Li_2022}, it appears likely that they were satellite dwarf companions orbiting the progenitor galaxy of LMS-1/Wukong \citep[also see][]{Malhan_2021}. This evidence suggests that LMS-1/Wukong originated from a particularly ancient and MP dwarf galaxy that formed at an early epoch of galaxy formation. According to dynamical analyses by \cite{Malhan_2021}, the LMS-1 progenitor was likely accreted onto the Milky Way at least $\gtrsim8-10$ Gyr ago.

In the action diamond (Fig. \ref{fig:Lz-E_action_diamond}, right), \object{HE2159-0551} is clearly located near stars identified as members of LMS-1/Wukong. This visualization emphasizes orbital characteristics such as radiality and angular momentum components, which are particularly useful for distinguishing accreted populations from stars formed in-situ within the Milky Way. Interestingly, \object{HE2159-0551} exhibits a distinctly prograde orbital signature compared to the main body of LMS-1/Wukong stars, suggesting a unique dynamical history. Such a prograde orbit typically characterizes stellar populations of the Galactic thick disk, indicating that \object{HE2159-0551} may represent a special case of a LMS-1/Wukong star that has dynamically migrated toward thick-disk orbital properties. We note that two additional MP stars have been discovered in an independent study by Monty et al. (in prep.), which also seem to belong to LMS-1/Wukong.

This scenario implies the existence of a previously unrecognized transition region or overlap zone between the LMS-1/Wukong merger debris and the thick-disk component of the Galaxy, potentially created through dynamical mixing processes such as disk heating or radial migration. Furthermore, the position of \object{HE2159-0551} in this orbital parameter space raises the question of possible overlaps or transitions between LMS-1/Wukong and other known substructures, notably the Helmi stream. To address this possibility, we apply the orbital definition of the Helmi stream from \citet{Koppelman_2019b}, namely $0.70 < L_z < 1 \, [10^3 \, \text{kpc km s}^{-1}]$ and $1.60 < L_\perp$ \footnote{With the definition of the perpendicular counterpart of $L_z$ with $L_\perp = \sqrt{L^2_x+L^2_y}$.} $< 3.20 \, \text{kpc km s}^{-1}$, to test for membership. Our analysis reveals that \object{HE2159-0551} does not fulfill these Helmi stream criteria. Thus, despite the apparent dynamical proximity, the star can be firmly excluded as a Helmi stream member. This reinforces the interpretation that \object{HE2159-0551} represents an intriguing transitional object situated in a kinematical overlap region connecting the LMS-1/Wukong merger event with the Milky Way’s thick disk, rather than bridging to the Helmi stream. This finding highlights the complexity of stellar orbital histories and the importance of precise chemodynamical analyses to fully disentangle Galactic substructures. Such findings reinforce the importance of LMS-1/Wukong as a crucial tracer for investigating galaxy formation and early nucleosynthesis.

\section{Chemical abundances}\label{section:abundances}
\begin{figure*}
	\includegraphics[width=\linewidth, height=8cm]{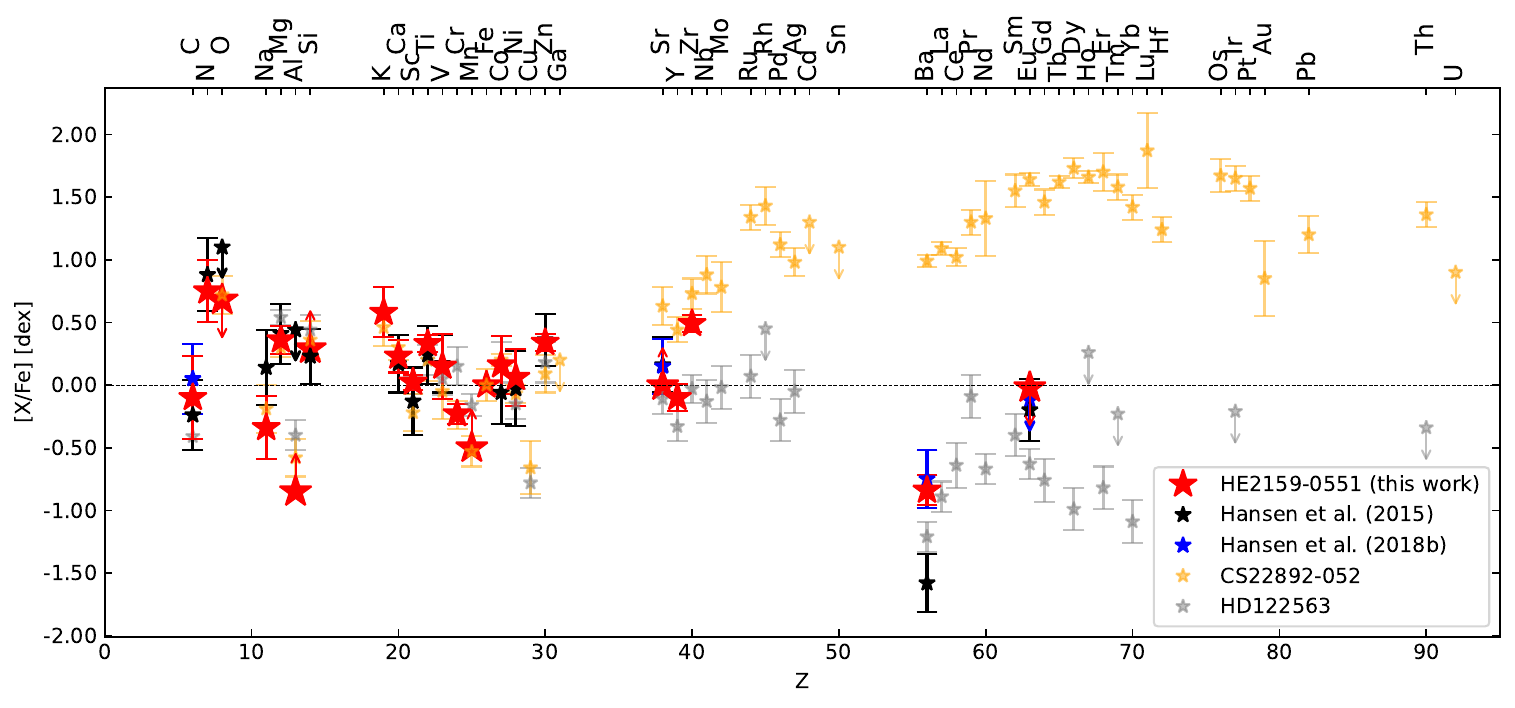}
	\caption{[X/Fe] versus atomic number Z for \object{HE2159-0551} (this work: red stars), with uncertainties. Data points with a downwards or upwards facing arrow are upper or lower limits, respectively. \object{HE2159-0551} is compared to \citet[black]{hansen2015elemental}, \citet[blue]{hansen2018r}, as well as CS22892-052 \citep[orange]{sneden2003extremely}, and HD122563 \citep[gray]{honda2004spectroscopic,honda2006neutron}. The horizontal dotted line shows [X/Fe] = 0.00 dex.}
	\label{fig:abundances}
\end{figure*}

\begin{figure*}
    \includegraphics[width=\linewidth]{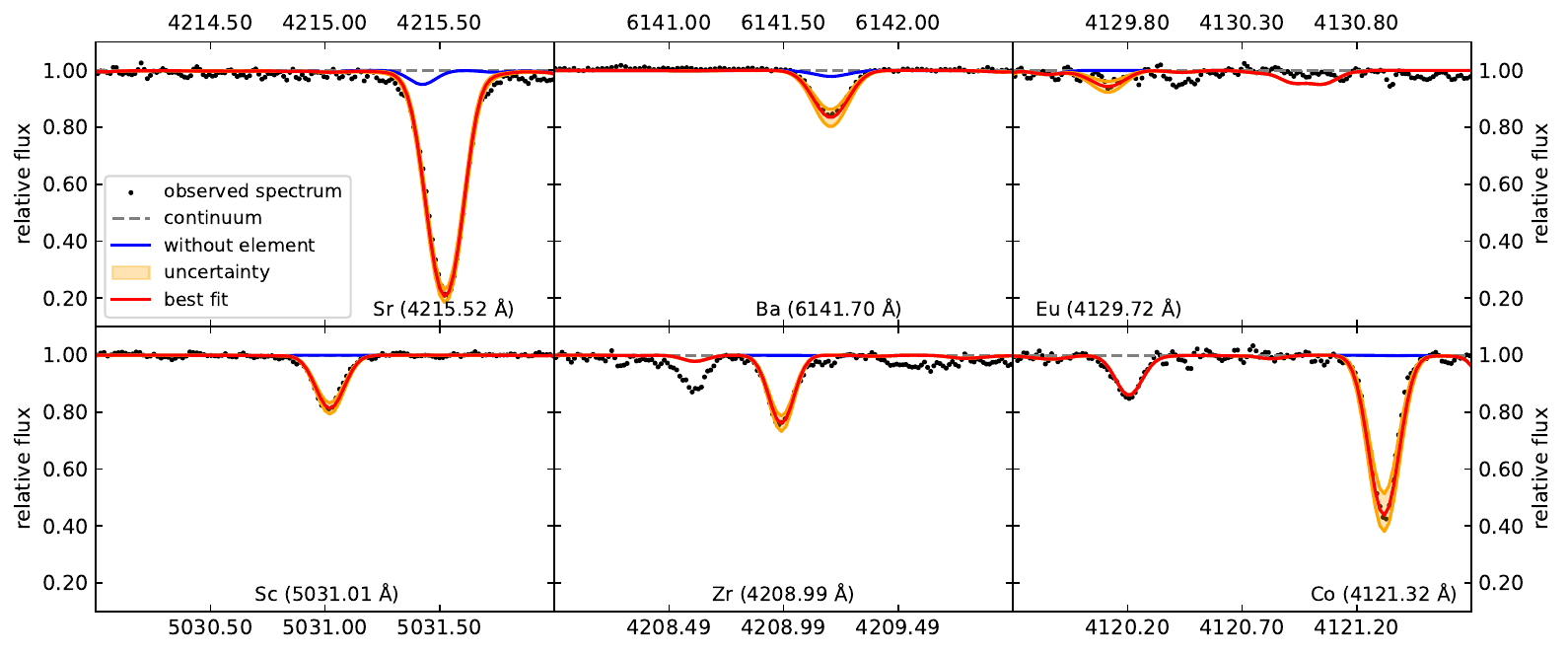}
	\caption{Top: Line synthesis of Sr, Ba, and Eu 
    with best fit indicated in red: [Sr/Fe] $\geq$ $-$0.01 dex (saturated), [Ba/Fe] = $-$0.84 dex, and [Eu/Fe] $\leq$ $-$0.02 dex (noisy). Bottom: [Sc/Fe] = 0.02 dex, [Zr/Fe] = 0.39 dex, and [Co/Fe] = 0.16 dex. Uncertainty shaded in yellow and observed spectrum in black. Continuum in gray and a blue line denotes the absence of the element measured.}
	\label{fig:abundance-spectra}
\end{figure*}

\begin{figure}
	\includegraphics[width=\linewidth]{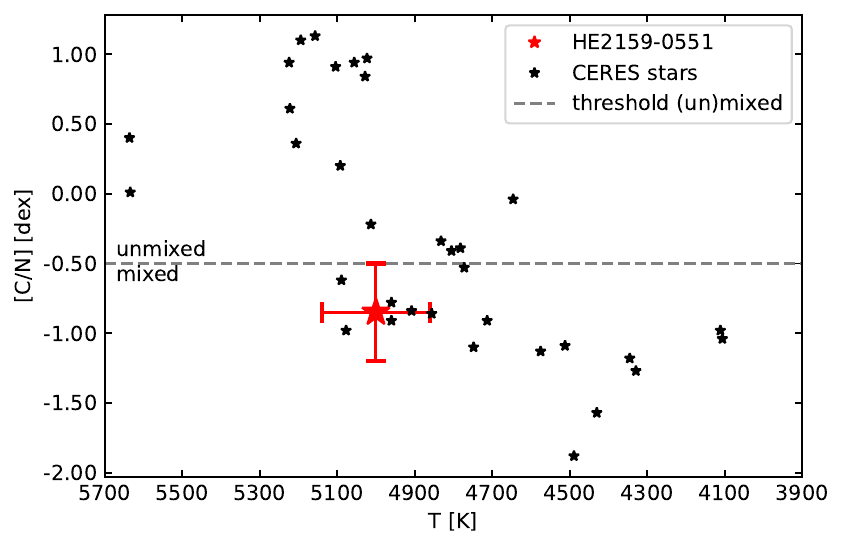}
	\caption{[C/N] ratio versus temperature of \object{HE2159-0551} (red) in comparison to the CERES project stars \citep[black]{FernandesdeMelo2024}, distinguishing between mixed (below the gray line) and unmixed (above the gray line) stars.}
	\label{fig:mixing}
\end{figure}

\begin{figure*}
    \includegraphics[width=\linewidth]{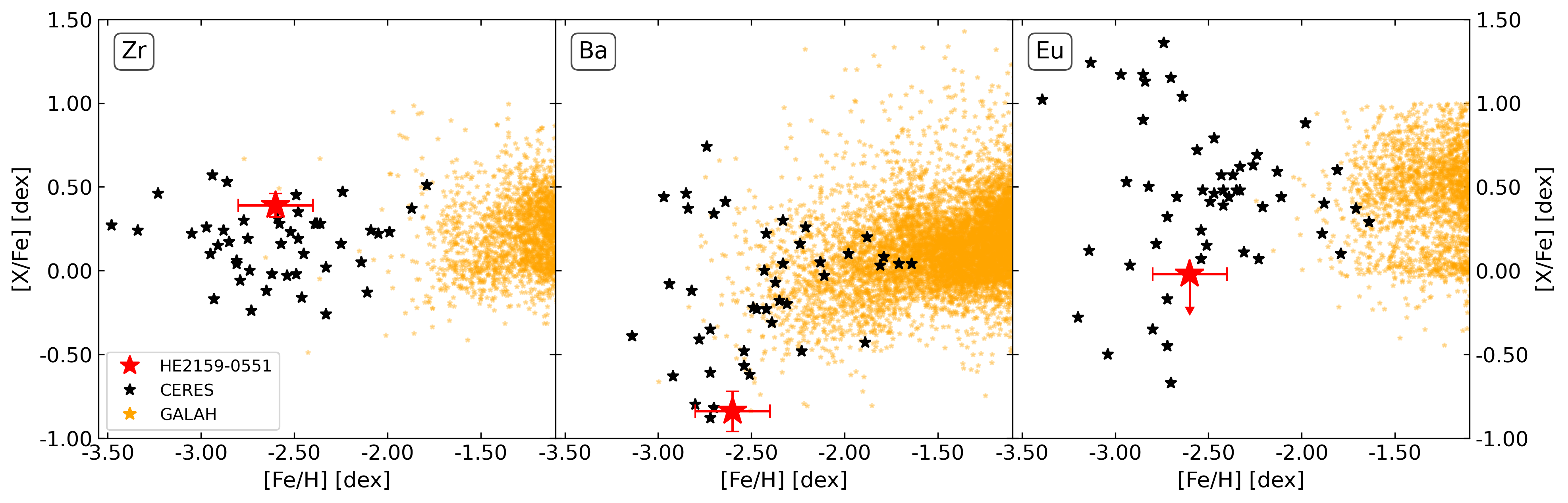}
     \caption{Comparison of \object{HE2159-0551} (this work, red), the CERES project sample \citep[black]{lombardo2022chemical,lombardo2025}, and the GALAH sample \citep[orange]{buder2024galahsurveydatarelease} we also use in Sect. \ref{section:kinematics}.}
    \label{fig:GALAH_CERES}
\end{figure*}

We utilize the software PyMoogi and its synthesis driver to derive chemical abundances or place limits (indicated as []) on 23 elements: C, N, [O], Na, Mg, [Al, Si], K, Ca, Sc, Ti, V, Cr, [Mn], Fe, Co, Ni, Zn, [Sr], Y, Zr, Ba, and [Eu] (see Table \ref{tab:abundances} and Figs. \ref{fig:abundances} to
\ref{fig:GALAH_CERES}). We conduct a visual comparison between the observed and synthetic spectra, striving to minimize the standard deviation to achieve the closest possible match between the two spectra. We determine the Fe abundance using EWs, ensuring that the spectral lines are clean. To validate our results, we following perform spectral synthesis of Fe. Line lists are generated using Linemake\footnote{\url{https://github.com/vmplacco/linemake}} \citep{placco2021linemake}.
We calculate the uncertainty for all elemental abundances by changing the stellar parameters one by one by their attributed uncertainty, while acknowledging that statistical errors may also arise from factors such as continuum placement. 

The derived chemical abundances are provided in Table \ref{tab:abundances} and are scaled to the solar abundances of \citet{asplund2009chemical}. In Fig. \ref{fig:abundances} we present all derived abundances and limits including the uncertainties of each individual element in comparison to the results for \object{HE2159-0551} by \cite{hansen2015elemental,hansen2018r}, CS22892-052 \citep{sneden2003extremely}, and HD122563 \citep{honda2006neutron,honda2004spectroscopic}. In Fig. \ref{fig:abundance-spectra} we provide examples of individual synthesized lines for six different elements (Sr, Ba, Eu, Sc, Zr, and Co). Figures \ref{fig:mixing} and \ref{fig:GALAH_CERES} provide comparisons of specific element abundances to literature data from the CERES (Chemical Evolution of R-process Elements in Stars) project \citep{lombardo2022chemical,lombardo2025,FernandesdeMelo2024} and the GALAH sample \citep{buder2024galahsurveydatarelease}.

We combine spectra from different observing nights to enhance the spectral SNR compared to earlier studies. The abundances derived in \citet{hansen2015elemental,hansen2018r} are listed in Table \ref{tab:abundances-lit-comparison} and illustrated in Fig. \ref{fig:abundances} together with the abundances from this study. A comparison of the two papers reveals small differences in their derived abundances, which is also reflected in their metallicity values: $-$2.81 dex in \citet{hansen2015elemental} and $-$2.75 dex in \citet{hansen2018r}. Overall, most abundances from the literature of \citet{hansen2015elemental,hansen2018r} agree with our results within combined uncertainties, indicating strong consistency and reliability across studies. However, different lines can be sensitive to varying conditions (e.g., Fe II lines being more gravity-sensitive than Fe I lines), leading to some discrepancies. A careful selection of lines is crucial and can account for discrepancies when comparing to literature values.

\begin{figure}
    \centering
    \includegraphics[width=\linewidth]{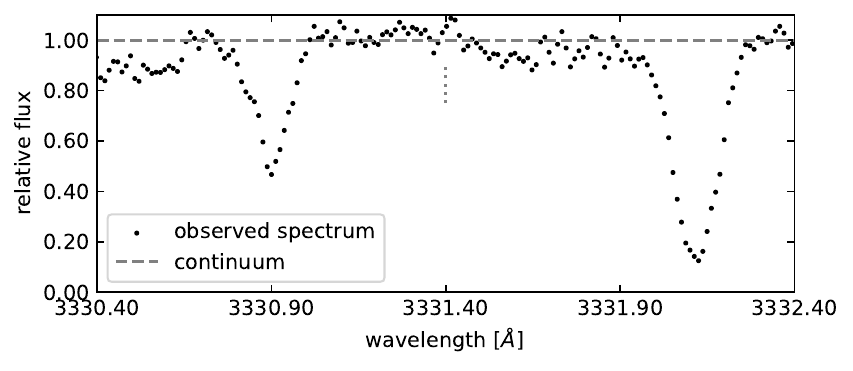}
    \caption{Non-detection of Gd. The perpendicular gray line indicates the wavelength of Gd at 3331.40 \AA{} and shows that no abundance/limit can be derived/placed for this element.}
    \label{fig:Gd-line-3331}
\end{figure}

We also attempt to derive abundances or place limits for other elements: Li, Be, S, Cu, Rb, Mo, Ru, Pd, Ag, La, Ce, Pr, Sm, Gd, Dy, Ho, Er, Os, Pt, Pb, Th, and U (see also Sect. \ref{section:discussion-conclusion}). Unfortunately, these attempts are unsuccessful. The primary reason is that the spectral features for these elements are extremely weak and in many cases no discernible line is present. The expected position of the lines do not exhibit any signal distinguishable from the spectral noise, precluding the possibility of reliably measuring abundances or even setting limits. The noise dominates the relevant spectral regions and any feature that might correspond to a heavy-element line is completely buried. This likely has its ties in the peculiar nucleosynthesis event that enriched the star partially and in too low levels of most heavy elements. As an illustrative example, Fig. \ref{fig:Gd-line-3331} shows the expected position of the Gd line in the blue region of the spectrum, highlighting the absence of a detectable line. This plot is representative of the spectral appearance of the heavy elements throughout the spectra.

Furthermore, we compare our seemingly r-peculiar star to two other well-known stars: CS22892-052 (r-rich; \citet{sneden2003extremely}) and HD122563 (r-poor; \citet{honda2006neutron,honda2004spectroscopic}) in Fig. \ref{fig:abundances}. Both studies provide a comprehensive range of abundances for the individual VMP stars. Notably, the abundances of heavier elements vary among the three stars. \object{HE2159-0551} aligns more closely with the star HD122563 considering that we find no detection for elements with Z $>$ 56.

\begin{table*}
    \caption{Overview of derived abundances and uncertainties ($\sigma$) organized by atomic number.}
	\centering
	\label{tab:abundances}
	\begin{tabular}{lllllll}
	\hline \hline
	   Element & [X/Fe] & [X/H] & $\sigma$ & Comment & Used lines & Reference\\
   & [dex] & [dex] & [dex] & & [\AA{}] &\\
         \hline
		C I& $-$0.10 & $-$2.70 & 0.33 &   -- & 4300 (CH band) & [1] \\
		N I &  $~~$0.75  & $-$2.00 & 0.25 &  --   & 3360 (NH band) & \\
        O I& $~~$0.68  &  $-$1.92 & 0.41 &  UL & 6300.30 & [2] \\ 
        Na I&  $-$0.34 &  $-$2.94 & 0.25 & -- & 8183.26, 8194.82& \\ 
		  Mg I&   $~~$0.36 & $-$2.24 & 0.11 & -- & 5528.40,  5711.08 & [3] \\
        Al I&   $-$0.85 & $-$3.45 & 0.44 & LL & 3961.52 & [4] \\
        Si I& $~~$0.29  & $-$2.31 & 0.19 & LL & 3905.52, 7034.90, 7918.38, 7944.00 & [5] \\
        K I & $~~$0.58 & $-2.02$ & 0.20 & -- & 7698.98 & \\
        Ca I &   $~~$0.23 & $-$2.43 & 0.13 & -- & 6102.72, 6162.17, 6169.06, 6169.56 & [6] \\
		  Sc II&  $~~$0.02 & $-$2.58 & 0.06 & --  & 5031.01, 5526.79  & [7] \\
		Ti II& $~~$0.33 & $-$2.27 & 0.07  & -- & 4708.67, 5185.91, 5226.55, 5336.79, 5381.02 & [8] \\
        V I/II & $~~$0.15 & $-$2.45 & 0.26  & -- & 3951.96, 4002.93, 4005.71, 4379.23 & [9] \\
        Cr I& $-$0.23 & $-$2.83 & 0.08 & -- & 5204.50, 5208.41, 5247.58 & [10] \\
        Mn I& $-$0.50 & $-$3.10  & 0.20    & LL & 4030.76, 4033.07, 4034.49 & [11] \\
		Fe I &  $~~$0.00 & $-$2.60 &  0.19 & -- &3443.87,  3475.45, 3490.57, 3815.84, 3840.43, 3841.04, 3856.37& \\
            & & & & & 3872.50, 3949.95, 4134.67, 4210.34,  4271.15, 4271.76,  4299.23  & \\
            & & & & &  4404.75,  4494.56, 4920.50, 5269.53, 5328.03, 5371.49 & [12] \\
        Fe II &  $~~$0.00 & $-$2.60 &  0.19 & -- & 4233.17, 4296.57, 4416.83, 4508.28, 4520.22,   4629.33,  4731.45  & \\
             & & & & & 4923.92 , 5316.61, 5534.84,  6247.55, 6432.68, 6456.38, 6516.08 & [13] \\
    	Co I& $~~$0.16 & $-$2.44 & 0.23 & -- &3845.47, 4121.32 & [14] \\
        Ni I  &  $~~$0.06 & $-$2.54 & 0.23 & -- & 7110.90, 7748.88, 7788.93, 7797.58 & \\
        Zn I & $~~$0.34 & $-$2.26 & 0.07 & -- & 4810.53 & [15] \\
		Sr II &   $-$0.01 & $-$2.61 & 0.20 & LL &4077.71, 4215.52  & [16] \\
		Y II&  $-$0.18 & $-$2.78 & 0.11 & -- & 3774.33, 5087.43, 5200.42, 5205.73 & [17] \\
		Zr II &   $~~$0.39 & $-$2.21 & 0.07 & -- & 3499.57, 4050.33, 4208.99 & [18] \\
	    Ba II &   $-$0.84 & $-$3.44 & 0.12 & -- & 5853.70, 6141.70, 6496.90 & \\		
		Eu II&  $-$0.02 & $-$2.62 & 0.16 & UL & 4129.72 & [19] \\		
		\hline
	\end{tabular}
    \tablefoot{The comment indicates an upper (UL) or lower limit (LL), due to weak lines or saturation. Lines used in this study are listed in column 6.}
    \tablebib{[1] \citet{masseron2014ch}; [2] \citet{magg2022observational}; [3] \citet{rhodin2017experimental}; [4] \citet{roederer2021detection}; [5] \citet{den2023atomic,rhodin2024accurate}; [6] \citet{den2021atomic}; [7] \citet{lawler2019transition}; [8] \citet{lawler2013improved}; [9] \citet{2014ApJS..214...18W,2014ApJS..215...20L,2016ApJS..224...35H,2018ApJS..234...25W}; [10] \citet{sobeck2007improved}; [11] \citet{den2011improved}; [12] \citet{belmonte2017fe,den2014fe,ruffoni2014fe}; [13] \citet{den2019atomic};  [14] \citet{wood2013improved,lawler2018transition}; [15] \citet{roederer2012detection,bergeson1993radiative}; [16] \url{https://www.nist.gov/pml/atomic-spectra-database}; [17] \citet{hannaford1982oscillator,biemont2011lifetime}; [18] \citet{ljung2006new,malcheva2006radiative}; 
      [19] \citet{lawler2001improved}
   }
\end{table*}

\begin{table}
	\caption{Abundances and uncertainties derived in this study compared to literature values of \citet[][(1)]{hansen2015elemental} and \citet[][(2)]{hansen2018r}.}
	\label{tab:abundances-lit-comparison}
	\begin{tabular}{lrrc}
        \hline \hline
		 Element & [X/Fe] (this work) & [X/Fe] (1) & [X/Fe] (2) \\
    & [dex]$~~~~~$ & [dex]$~~~~~$  & [dex]  \\
		\hline
		C & $-$0.10 $\pm$ 0.33 & $-$0.24 $\pm$ 0.28 &  0.05 \\
		N & 0.75 $\pm$ 0.25 & 0.88 $\pm$ 0.29 & -- \\ 
        O & $\leq$ 0.68 $\pm$ 0.41 & < 1.10 & -- \\
        Na & $-$0.34 $\pm$ 0.25 & 0.14 $\pm$ 0.30 & -- \\
		Mg & 0.36 $\pm$ 0.11 & 0.41 $\pm$ 0.24 & --\\
        Al & $\geq$ $-$0.85 $\pm$ 0.44 & < 0.44 & -- \\
        Si & $\geq$ 0.29 $\pm$ 0.19 & 0.23 $\pm$ 0.22 & -- \\
        Ca & 0.23 $\pm$ 0.13 & 0.17 $\pm$ 0.23 & -- \\
		Sc &  0.02 $\pm$ 0.06 & $-$0.13 $\pm$ 0.27 & --\\
		Ti & 0.33 $\pm$ 0.07 & 0.24 $\pm$ 0.23 & -- \\
        V & 0.15 $\pm$ 0.26 & 0.17 $\pm$ 0.23 & -- \\
        Cr & $-$0.23 $\pm$ 0.08 & $-$0.40 $\pm$ 0.24 & -- \\
        Mn & $\geq$ $-$0.50 $\pm$ 0.20 & $-$0.45 $\pm$ 0.25 & -- \\ 
        $[$Fe/H$]$ & $-$2.60 $\pm$ 0.20 & $-$2.81 $\pm$ 0.20 & $-$2.75 $\pm$ 0.18 \\
		Co  & 0.16 $\pm$ 0.23 & $-$0.06 $\pm$ 0.25 & -- \\
        Ni & 0.06 $\pm$ 0.23 & $-$0.03 $\pm$ 0.30 & -- \\
        Zn & 0.34 $\pm$ 0.07 & 0.36 $\pm$ 0.21 & -- \\
		Sr &  $\geq$ $-$0.01 $\pm$ 0.10 & 0.16 $\pm$ 0.22 &  0.15 \\
		Ba &   $-$0.84 $\pm$ 0.20 & $-$1.58 $\pm$ 0.23 & $-$0.75 \\
		Eu &  $\leq$ $-$0.02 $\pm$ 0.16 & $-$0.20 $\pm$ 0.25 & < $-$0.12 \\	
		\hline
	\end{tabular}
\end{table}

\subsection{Carbon, nitrogen, and oxygen}
We derive abundances for C and N using the CH and NH molecular bands, yielding [C/Fe] = $-$0.10 $\pm$ 0.33 dex and [N/Fe] = 0.75 $\pm$ 0.25 dex. The star is N-enhanced, but not C-enhanced, following the criteria of \citet{beers2005discovery} and \citet{aoki2007carbon}, who define carbon-enhanced MP (CEMP) stars as [C/Fe] > $+$0.70 dex. The measured [C/N] = $-$0.85 dex indicates that mixing has occurred, see Fig. \ref{fig:mixing}. Hence, mixing processes have affected the star's lighter element abundances \citep[e.g.,][]{spite2005first}. \citet{placco2014carbon} provide a correction\footnote{\url{https://vplacco.pythonanywhere.com/}} accounting for C depletion due to mixing. Our [N/H] deviates from solar, rendering these corrections inaccurate. 
We find [O/Fe] $\leq$ 0.68 $\pm$ 0.41 dex. Our C and N abundances agree within uncertainties when comparing to \citet{hansen2015elemental}, who use molecular bands around 4300 \AA, 3360 \AA, 3890 \AA, and 4215 \AA{} -- mostly consistent with those used in this work, see Table \ref{tab:abundances}.

\subsection{Sodium, aluminum, chromium, scandium, and manganese}
We determine [Na/Fe] = $-$0.34 $\pm$ 0.25 dex. We also inspect the Na lines at 5889.95 \AA{} and 5895.95 \AA{}. Due to saturation issues, we neglect the abundance of those lines when determining the final mean Na abundance value. The Al line yields a lower limit (LL) of [Al/Fe] $\geq$ $-$0.85 $\pm$ 0.44 dex as we had to rely on a blended, saturated line. While \citet{baumuller1997aluminium} report near-solar [Al/Fe] ratios in VMP stars, our value is lower, but falls within the observed scatter around [Fe/H] $\approx$ $-$2.60 dex, which they attribute to NLTE effects. We find [Cr/Fe] = $-$0.23 $\pm$ 0.08 dex, consistent with the MP average in \citet{franccois2004evolution}. We determine [Sc/Fe] = 0.02 $\pm$ 0.06 dex, which shows less enhancement, consistent with expectations for MP stars \citep{cayrel2004first,frebel2008metal,hayek2009hamburg,hansen2011diss}. For Mn, we find a limit of [Mn/Fe] $\geq$ $-$0.50 $\pm$ 0.20 dex from blended, but also saturated lines. Despite the large uncertainty, the result is in line with the studies by \citet{gratton1989abundance} and \citet{sobeck2006manganese}, who report subsolar Mn abundances decreasing with decreasing metallicity. However, near [Fe/H] $\approx$ $-$2.00 dex, they observe an increase in the Mn abundance. \object{HE2159-0551} seems to follow this trend. To probe the star's origin, we compute [Mg/Mn] $\leq$ 0.86 dex, see Sect. \ref{sect:abundance-planes}.

\subsection{Magnesium, silicon, calcium, and titanium}
We derive [Mg/Fe] = 0.36 $\pm$ 0.11 dex, [Ca/Fe] = 0.23 $\pm$ 0.13 dex, and [Ti/Fe] = 0.33 $\pm$ 0.07 dex. We place a LL of [Si/Fe] = $\geq$ 0.29 $\pm$ 0.23 dex due to saturation issues. The Mg, Si, and Ti enhancements ($\approx$ 0.35 dex) are typical for VMP stars and reflect early $\alpha$-element enrichment from Type II SNe during the galaxy's early chemical evolution. Our derived Mg, Si, and Ti abundances are consistent within uncertainties when comparing with \citet{hansen2015elemental}. The Ca abundance is slightly higher but still compatible. This is in line with the Ca values derived in Sagittarius (Sgr) in VMP and extremely MP stars \citep{Hansen2018Sgr}.

Based on [C/Mg] = $-$0.46 dex, we classify the star as multi-enriched, implying enrichment from multiple nucleosynthetic events \citep{hartwig2018descendants}. This is consistent with our expectations, although such classifications are subject to uncertainties from gas dilution, mixing, and variations in progenitor SNe properties \citep{hansen2020mono}.

\subsection{Potassium, vanadium, cobalt, nickel, and zinc}
We derive [K/Fe] = 0.58 $\pm$ 0.20 dex, [V/Fe] = 0.15 $\pm$ 0.26 dex, [Co/Fe] = 0.16 $\pm$ 0.23 dex, [Ni/Fe] = 0.06 $\pm$ 0.23 dex, and [Zn/Fe] = 0.34 $\pm$ 0.07 dex, consistent with results of other stars in \citet{da2018oxygen}, who report Zn behaving similarly to $\alpha$-elements. While Zn shows scatter near solar metallicity \citep{duffau2017gaia}, elevated [Zn/Fe] ratios are attributed to supernova (SN) enrichment \citep[][and references therein]{da2018oxygen}. At [Fe/H] $\approx$ $-$3.00 dex, Zn abundances in the Milky Way halo can reach supersolar abundances of [Zn/Fe] $\approx$ 0.50 dex \citep{duffau2017gaia}, placing our result within the expected range given the uncertainties. The abundance results of this subsection agree within combined uncertainties when compared to \citet{hansen2015elemental}.

\subsection{Strontium, yttrium, zirconium, and barium}\label{sec:abundances-Sr-Y-Zr-Ba}
We attempt to determine the Sr abundance using the 4077.71 \AA{} and 4215.52 \AA{} lines. However, due to the strong and saturated lines we report a LL of [Sr/Fe] $\geq$ $-$0.01 $\pm$ 0.20 dex. The weaker Sr line at 4607.33 \AA {} is not covered by our spectra. Although we use the same lines as \citet{hansen2015elemental}, our derived abundance is lower. In contrast, \citet{hansen2018r} primarily use the 4215.52 \AA {} line due to similar saturation issues. 
\citet{hansen2012silver} report a mean abundance of [Sr/Fe] $\approx$ 0.14 dex down to [Fe/H] = $-$2.50 dex in a larger MP sample, with increased scatter at lower metallicities -- our result lies well within this scatter.

We derive [Y/Fe] = $-$0.18 $\pm$ 0.11 dex, [Zr/Fe] = 0.39 $\pm$ 0.07 dex, and [Ba/Fe] = $-$0.84 $\pm$ 0.12 dex. The spread of more than 1.20 dex between these element abundances is notable. Y and Zr are primarily s-process elements, with contributions of 77.80 \% and 81.70 \%, respectively \citep{prantzos2020chemical}\footnote{In this and all following subsections, we reference \citet{prantzos2020chemical} to provide the production fractions for the s- and r-processes of individual elements for understanding their formation, nucleosynthetic history, and the broader context of GCE.}; however, at low metallicities around [Fe/H] = $-$2.40 dex and below, the production of these elements is influenced by the limited presence of AGB stars, as well as contributions from other stellar processes \citep{simmerer2004rise,lombardo2025}. \citet{hansen2012silver} suggest Zr may originate from both the weak r- and main s-process, reporting a mean [Zr/Fe] $\approx$ 0.20 dex for [Fe/H] $\geq$ $-$2.50 dex, but with increasing scatter at lower metallicities; consistent with our results. The Y and Zr abundances agree with those in \citet{mello2014high}, \citet{lombardo2022chemical}, and \citet{franccois2007first} for stars of similar metallicity. The elevated Zr and low Y might be explained by SN enrichment or the $\nu$p-process, as internal stellar mixing appears to have little or no effect on Sr, Y, and Zr abundances \citep{FernandesdeMelo2024}.

Barium is mainly formed by the s-process in AGB stars. The low Ba abundance combined with the solar scaled Y, however, makes AGB stars an unlikely formation site, see also Sect. \ref{section:nucleosynthesis-models} in the appendix. Our Ba abundance follows the trend of MP stars \citep{lombardo2022chemical,lombardo2025}, see Fig. \ref{fig:GALAH_CERES}. 
The large Ba scatter at [Fe/H] < $-$ 2.50 dex, also seen in the sample of \citet{hansen2012silver} and \citet{franccois2007first}, may reflect poor interstellar medium (ISM) mixing or cosmic variations \citep{barklem2005hamburg,franccois2007first}, warranting further investigation with additional sample stars. Ba abundances of \object{HE2159-0551} vary across studies of \citet{hansen2015elemental} and \citet{hansen2018r}, but our results agree well with \citet{hansen2018r}, who use the same lines as we do (5853.70 \AA, 6141.70 \AA, and 6496.90 \AA). This low Ba abundances agrees with the weak r- or $\nu$p-process forming Sr-Zr but not reaching Ba. This could be associated with an intermediate/massive SN producing Ca to Zr, but not Ba or heavier. We speculate that the low Ba abundance we find in this star might be attributed to the LMS-1/Wukong merger event, as the kinematical analysis of this star indicates that it originated from this merger event, see Sect. \ref{subsec:kinematics-origins}. This highlights the significance for investigating stars chemodynamically and taking all properties into consideration. The low Ba abundance for LMS-1 candidates has been derived in two other stars in an independent study by Monty et al. (in prep.) and may be a tracer for LMS-1.

\subsection{Heavy element upper limits}
We place an UL of [Eu/Fe] $\leq$ $-$0.02 $\pm$ 0.16 dex, based on a weak line affected by significant noise. As Eu is 95.10 \% r-process produced, the low abundance is noteworthy for a VMP star. Compared to the CERES stars \citep{lombardo2025} and the GALAH sample \citep{buder2024galahsurveydatarelease}, our target only marginally follows the trend of decreasing Eu with metallicity and lies outside the main distribution in Fig. \ref{fig:GALAH_CERES}, suggesting a lack of r-process and overall n-capture enhancement. The result is consistent with \citet{hansen2015elemental,hansen2018r}. While surface abundances may be affected by mixing, elements like Sr, Y, and Zr are not expected to show such noteworthy effects \citep{FernandesdeMelo2024}, supporting the interpretation that [Ba/Eu] reflects the star’s initial composition. 

Based on the heavy element abundances/limits we find no strong s- or r-process enhancement. A kind of weak n-capture or $\nu$p-process may have enriched this star, which explains the elemental abundances up to Zr and little Ba. The heavier limit supports the result of (multi-)enrichment of the ISM mixed poorly into an ISM with a low abundance of Ba and Eu, agreeing with multi-enrichment, likely the weak r-/$\nu$p-process following a SN II.

\subsection{Abundance planes}\label{sect:abundance-planes}
Our goal is to determine whether \object{HE2159-0551} formed in-situ or was accreted into the Galaxy. In principle, abundance planes such as [Mg/Fe] vs. [Fe/H], [C+N/Fe] vs. [Fe/H], [Al/Fe] vs. [Fe/H], and particularly [Mg/Mn] vs. [Al/Fe] have been successfully used to distinguish between accreted and in-situ stellar populations \citep{Freeman_Bland_Hawthorn_2002,Tolstoy_2009,Hawkings2015,Das2020}. [Mg/Mn] is used to highlight the $\alpha$-poor population rather than [Mg/Fe] as Mn is a pristine tracer of SN Ia. Moreover, since SN Ia do not produce Al as efficiently as Fe, the [Al/Fe] plane demonstrates to distinguish between accreted stars born in less massive systems, whereas in-situ stars display elevated [Al/Fe] and lower [Mg/Mn] values.

Unfortunately, in the case of \object{HE2159-0551}, we can only place limits for abundances of key diagnostic elements such as Al and Mn. Due to these constraints, the star’s position in these abundance diagrams is uncertain and provides only limited diagnostic power. Although our placed UL on [Al/Fe] and LL on [Mg/Mn] would, in principle, support an accreted origin, the lack of precise abundances makes such conclusions tentative. For this reason, we refrain from explicitly presenting these abundance planes. Nonetheless, we emphasize that we carefully inspected these chemical signatures and find that the overall chemodynamical evidence, particularly the orbital parameters consistent with LMS-1 membership, remains compelling. Thus, despite the limited diagnostic value from abundances of Al and Mn, we maintain our interpretation of \object{HE2159-0551} as an exceptional LMS-1 star, dynamically mixed into a thick-disk region of the Galaxy.

\section{Discussion and conclusion}\label{section:discussion-conclusion}
We present kinematics and more abundance measurements for the star \object{HE2159-0551} than previously reported in the literature, revealing a peculiar abundance pattern and a connection to the LMS-1/Wukong merger event. Our results show an overall good agreement with the stellar parameters and abundances of \citet{hansen2015elemental,hansen2018r}. Accurate determination of these parameters is crucial, as uncertainties in stellar parameters can influence abundance results by several tenths of a dex.

By utilizing spectra collected over several years, we are able to improve both the reliability and SNR of our data, as well as refute binarity. 
A complete set of clean lines and accurate atomic physics (excitation potential and oscillator strengths) ensure that the abundance estimates are not biased by noise or uncertainties in individual measurements.  

In addition to the elements mentioned above, we also attempt to derive abundances for Li, Be, S, Cu, Rb, Mo, Ru, Pd, Ag, La, Ce, Pr, Sm, Gd, Dy, Ho, Er, Os, Pt, Pb, Th, and U. Unfortunately, the abundances of these elements cannot be determined due to the presence of very weak or non-existent spectral lines (see Sect. \ref{section:abundances}), in the spectra with SNR of about 70 at 4000 \AA{} and 100 at 5000 \AA. The inability to measure Li is particularly notable, as it suggest significant depletion of Li in this VMP star, consistent with the findings of \citet{korn2006probable}, confirming internal mixing. The internal mixing processes related to the red giant branch bump alter the surface abundances of certain lighter elements, but cannot explain the anomalies observed in the heavier elements \citep[][and references therein]{FernandesdeMelo2024}. The low Ba reflects more complex nucleosynthesis processes at play prior to the star's formation. 

The scatter observed in the abundances of elements such as Sr, Y, Zr, Ba, and Eu at [Fe/H] < $-$2.50 dex is often interpreted as a consequence of poor mixing in the ISM due to a limited number of SNe \citep{hansen2012silver,hansen2014many}. This scatter implies that the star’s chemical composition was shaped by only a few nucleosynthetic events, reinforcing the idea that \object{HE2159-0551} may have experienced an unusual enrichment history. A key feature of this star is its low abundances of n-capture elements, including Y and especially Ba. This suggests that the star has not undergone a third dredge-up event, nor has it been pre-enriched in s-process elements, distinguishing it from many other stars of similar metallicity. This unique characteristic points to a potentially distinct nucleosynthesis history for the star, which may be connected to the LMS-1/Wukong merger. This galaxy might exhibit a peculiar and generally low Ba abundance at low metallicity as chemical tracers.

In typical r-process-enhanced VMP stars, higher element abundances tend to resemble solar values \citep{barklem2005hamburg}; however, this star’s deviation suggest a completely different nucleosynthetic history. Numerous studies in the literature have identified stars exhibiting exceptionally low abundances of heavy elements \citep[e.g.,][]{koch2008highly,koch2013spectroscopic,fulbright2004draco}. In each of these cases, the prevailing interpretation suggests that the observed enrichment of heavy elements is primarily the result of contributions from a limited number of very massive stars. 
Considering that the star \object{HE2159-0551} is associated with the LMS-1/Wukong merger event, it is reasonable to attribute the divergent nucleosynthesis processes to this connection. 
The weak r- or $\nu$p-process in a supernova explosion may have contributed to the early ISM of LMS-1/Wukong, but the conditions were too proton (or neutrino) rich to efficiently form elements much heavier than Zr. Consequently, there is no strong r-enrichment from NSMs at [Fe/H] $\sim$ $-$2.50 dex; otherwise, we would expect a higher abundance level of Eu in this star.

The LMS-1/Wukong system was probably enriched primarily by a few normal, low-mass CC SN, which could facilitate e.g., a $\nu$p-process that forms Sr-Zr, but does not produce Ba in significant quantities more than 10 Gyr ago -- prior to the merger event. 
The galaxy merger event may have triggered additional SN explosions and star formation. The material from these SNe would have mixed within the LMS-1/Wukong system, which exhibits only a low level of r-process elements, possibly originating from the mini-halo stages. It is only above [Fe/H] $\sim$ $-$2.50 dex that NSMs would significantly enrich the LMS-1/Wukong system. This observations provides a promising constraint on the compact merger delay times, placing the LMS-1/Wukong system as a counterpart to the ultra-faint dwarf galaxy Reticulum II, which exhibits high levels of heavy n-capture abundances at low metallicities \citep{ji2016complete}. 

\cite{Jean-Baptist_2017} showed that a single accretion event can lead to multiple overdensities in orbital space \citep[see also][]{Koppelmann_2020}. An association between LMS-1 and Wukong seems plausible due to the close proximity between these two substructures in orbital planes (see Fig. \ref{fig:Lz-E_action_diamond}) and the chemical similarity, in particular the low Ba abundance we and Monty et al. (in prep.) find individually (see Sect. \ref{sec:abundances-Sr-Y-Zr-Ba}). It could suggest that they are indeed the same system, which has peculiar abundances at low metallicity. However, we defer providing a definitive conclusion until more elemental abundances are obtained in a larger sample of both LMS-1 and Wukong candidates.

\begin{acknowledgements}
      Based on observations collected at the European Organisation for Astronomical Research in the Southern Hemisphere under ESO programmes 095.D-0402(A) and 077.D-0035(A) and on data obtained from the ESO Science Archive Facility. 
      This work has made use of data from the European Space Agency (ESA) mission {\it Gaia} (\url{https://www.cosmos.esa.int/gaia}), processed by the {\it Gaia} Data Processing and Analysis Consortium (DPAC, \url{https://www.cosmos.esa.int/web/gaia/dpac/consortium}). Funding for the DPAC has been provided by national institutions, in particular the institutions participating in the {\it Gaia} Multilateral Agreement. 
      NOIRLab IRAF is distributed by the Community Science and Data Center at NSF NOIRLab, which is managed by the Association of Universities for Research in Astronomy (AURA) under a cooperative agreement with the U.S. National Science Foundation. 
      AS acknowledges the support by the State of Hesse within the Research Cluster ELEMENTS (Project ID 500/10.006). 
      AS was supported by the Deutsche Forschungsgemeinschaft (DFG, German Research Foundation) – Project-ID 279384907 – SFB 1245. 
      CJH acknowledges ChETEC-INFRA (EU project no. 101008324) and HFHF.  
      AS acknowledges the helpful input of Linda Lombardo, Raphaela Fernandes de Melo, Arthur Alencastro Puls, Alexander Dimoff, Andrew Gallagher, and Andreas Koch-Hansen. 
      We thank the referee for constructive feedback. 
      Finally, we acknowledge fruitful discussions with S. Monty and J. Bovy.   
      Software: Astropy \citep{astropy:2013,astropy:2018,astropy:2022}, Numpy \citep{harris2020array}, SciPy \citep{2020SciPy-NMeth}, Matplotlib \citep{Hunter:2007}, Galpy \citep{Bovy_2015}. 
\end{acknowledgements}

\bibliographystyle{aa} 
\bibliography{bibliography.bib}

\begin{appendix}

\section{Radial velocity}\label{sec:RV-appendix}
In Table \ref{tab:rvs} we present a detailed overview of the computed RVs discussed in Sect. \ref{section:radial-velocity}.

\begin{table*}
\caption{Computed RV values (single-file (shifting) and cross-correlation (CC) method) including computed barycentric correction (BC), separated by individual observations with information on exposure time, slit width, central wavelength, and SNR for each file.}
\label{tab:rvs}
\begin{tabular}{lllllllll}
\hline\hline
Date & Time & Exp. time & Slit width & Central wavel. & SNR & RV\textsubscript{BC} & RV\textsubscript{final, shifting} & RV\textsubscript{final, CC} \\
& [hh:mm:ss] & [s] &  [arcsec] & [nm] &  [px$^{-1}$] & [km\,s$^{-1}$] & [km\,s$^{-1}$] &  [km\,s$^{-1}$] \\
\hline
2006-04-19 & 09:07:29.889 & 900.00 & 0.7 & 580 & 78\tablefootmark{a} & 25.21 & $-$106.00 & $-$109.00 \\
2006-04-19 & 09:07:29.889 & 900.00 & 0.7 & 580 & 101\tablefootmark{b} &  25.21 & $-$107.00 & $-$106.00 \\
2006-04-19 & 09:07:35.649 & 900.00 & 0.8 & 390 & 11\tablefootmark{c} &  25.21 & $-$106.75 & $-$106.04 \\
2015-06-18 & 07:43:30.885 & 1492.00 & 0.7 & 346 & 15\tablefootmark{c} & 26.23 & $-$105.30 & $-$105.50 \\
2015-06-18 & 07:43:22.580 & 1482.00 & 0.7 & 760 & 178\tablefootmark{b} &  26.26 & $-$106.40 & $-$106.77 \\
2015-06-18 & 07:43:22.580 & 1482.00 & 0.7 & 760 & 239\tablefootmark{d} & 26.25 & $-$105.60 & $-$105.90 \\
2015-07-20 & 09:43:22.376 & 1218.00 & 0.7 & 390 & 13\tablefootmark{c} &  15.59 & $-$106.13 & $-$105.90 \\
2015-07-20 & 09:43:18.398 & 1208.00 & 0.7 & 564 & 124\tablefootmark{a} &  15.70 & $-$109.60 & $-$109.50 \\
2015-07-20 & 09:43:18.398 & 1208.00 & 0.7 & 564 & 185\tablefootmark{b} & 15.70 & $-$107.10 & $-$105.90 \\
2015-07-28 & 06:26:36.382 & 1492.00 & 0.7 & 346 & 32\tablefootmark{c} &  12.59 & $-$105.70 & $-$104.90 \\
2015-07-28 & 06:26:31.885 & 1482.00 & 0.7 & 760 & 203\tablefootmark{b}  & 12.59 & $-$104.80 & --\tablefootmark{e}\\
2015-07-28 & 06:26:31.885 & 1482.00 & 0.7 & 760 & 258\tablefootmark{d}  & 12.59 & $-$104.84 & $-$117.36\tablefootmark{f} \\
2015-08-23 & 04:25:58.245 & 1492.00 & 0.7 & 346 & 36\tablefootmark{c} &  0.22 & $-$105.80 & $-$104.90 \\
2015-08-23 & 04:25:56.548 & 1482.00 & 0.7 & 564 & 173\tablefootmark{a} &  0.22 & $-$107.40 & $-$109.50 \\
2015-08-23 & 04:25:56.548 & 1482.00 & 0.7 & 564 & 161\tablefootmark{b}& 0.22 & $-$107.00 & $-$106.00 \\
\hline	
 & & & & & & & $-$106.40 $\pm$ 1.20 & $-$107.00 $\pm$ 4.00\\
\hline		
\end{tabular}
\tablefoot{We measure the SNR for the files around \tablefoottext{a}{5010 \AA,}
\tablefoottext{b}{6010 \AA,} \tablefoottext{c}{3440 \AA, and} \tablefoottext{d}{8570 \AA, respectively.} \tablefoottext{e}{This file, which does not state a value for RV\textsubscript{final, CC}, is used as a template file.} \tablefoottext{f}{The last RV\textsubscript{final, CC} value of July 28, 2015 shows a bigger RV difference compared to the other values of that day, see Sect. \ref{section:radial-velocity}.} }
\end{table*}

\section{Stellar parameters -- uncertainties}\label{sec:StelllarParametersUncertanties}
We use the following assumptions for the computation of the uncertainties of the stellar parameters: we compute the effective temperature as 5000 K, with additional models $\pm$ 100 K, while keeping the other stellar parameters fixed. Similarly, we assess models for the surface gravity; the final value of 1.80 dex $\pm$ 0.20 dex. For the microturbulence, with a computed value of 2.40 km\,s$^{-1}$, we create models with$\pm$ 0.20 km\,s$^{-1}$. Using the outputs from these models, we compute uncertainties by analyzing the slope changes of the Fe I lines in response to variations in the individual parameters. We then quantify the uncertainty by computing the difference between the original value and the values from the modified models. We present the computation for $\log$ g here and apply the same procedure to the other two stellar parameters:
\begin{align}
	\Delta S_1 &= (\text{slope $\log$ g\textsubscript{1.60 dex}}) - (\text{slope $\log$ g\textsubscript{1.80 dex}}) \\
    \Delta S_2 &= (\text{slope $\log$ g\textsubscript{2.00 dex}}) - (\text{slope $\log$ g\textsubscript{1.80 dex}}).
\end{align}
We compute the average difference in slopes by calculating:
\begin{equation}
	\text{average slope difference} = \frac{|\Delta S_1| + |\Delta S_2| }{2}.	
\end{equation}
This leads to the uncertainty of e.g., the surface gravity:
\begin{equation}
	\Delta \log \text{g} = \frac{\text{average slope difference}}{\text{absolute value of the slope of best fit}}.
\end{equation}
With this, we get an uncertainty of 0.40 dex for $\log$ g. Following the same procedure for v\textsubscript{t} leads to an uncertainty of 0.30 km\,s$^{-1}$.

For the uncertainty in temperature, the slope change is a critical factor. A slope change of 0.001 for the Fe I lines corresponds to a variation in effective temperature of 3.22 K when the slope increases, and 2.32 K when it decreases. Thus, for simplicity we assume that a slope change of 0.001 results in an approximate temperature variation of 3 K. This allows us to compute the final uncertainty in temperature by considering the average slope difference for $\log$ g = 0.03, for v\textsubscript{t} = 0.02, and for T\textsubscript{eff} = 0.03 when comparing the computed models to the modified models. We find the uncertainty for T\textsubscript{eff} and $\log$ g as 90 K\footnote{Computed as 3000 K/slope $\cdot$ 0.03 = 90 K.} and for v\textsubscript{t} as 60 K\footnote{Computed as 3000 K/slope $\cdot$ 0.02 = 60 K.} by using the relation 3 K/0.001 = 3000 K/slope. Therefore, the uncertainty in the effective temperature is
\begin{align}
	\Delta T & =
	\sqrt{(\text{T\textsubscript{slope}})^2 + (\log \text{g\textsubscript{slope}})^2 + (\text{v\textsubscript{t,slope}})^2} \label{equ:effect-temp-uncert} \\
	& =  \sqrt{(90\: \si{\kelvin})^2 + (90\: \si{\kelvin})^2 + (60 \: \si{\kelvin})^2 } 
	&= 140\: \si{\kelvin}.
\end{align}

The actual temperature uncertainty could be expressed as $\pm$ 90 K (rounded to $\pm$ 100 K). However, this estimate does not account for the interdependencies among the stellar parameters. To incorporate these dependencies, we refer to equation \ref{equ:effect-temp-uncert} and report the final uncertainty value for the effective temperature as $\pm$ 140 K.

\section{Kinematics}\label{sec:appendix_kinematics}
We list the 6D phase-space values and their uncertainties, used in Sect. \ref{section:kinematics}, in Table \ref{tab:6d space-phase values}. 

\begin{table}[H]
    \centering
    \caption{6D phase-space values and their uncertainties  
    for \object{HE2159-0551}'s orbit and orbital parameters.}
    \begin{tabular}{lll}
    \hline \hline
         Parameter & Unit & \object{HE2159-0551} \\
         \hline
         \textit{Gaia} DR3 & & 2668962496722780544 \\
         $\alpha$ & [\degr] & 330.57\\
         $\delta$ & [\degr] & $-$5.61\\
         $D_\odot$ & [kpc] & 3.82 $\pm$ 0.20\\
         $\mu^*_\alpha$ & [mas/yr] & 0.92 $\pm$ 0.02\\
         $\mu_\delta$ & [mas/yr] & $-$4.36 $\pm$ 0.02\\
         RV & [km\,s$^{-1}$]& $-$106.40 $\pm$ 1.20\\
    \hline
    \end{tabular}
    \label{tab:6d space-phase values}
\end{table}

\section{Nucleosynthesis models}\label{section:nucleosynthesis-models}
\begin{figure*}
	\centering
	\includegraphics[width=\linewidth]{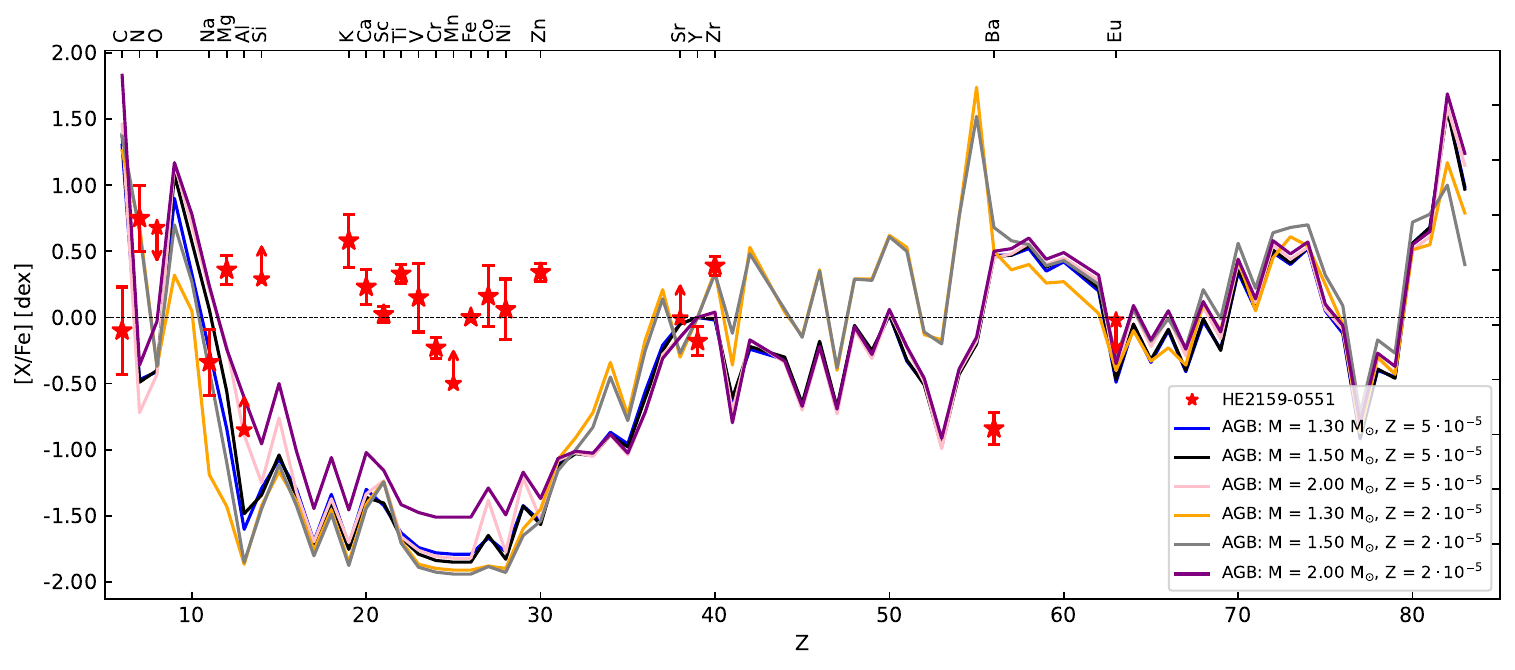}
    \caption{F.R.U.I.T.Y. models \citep{cristallo2011evolution} in comparison to the derived abundances of \object{HE2159-0551} (red stars). The upper/lower limits are represented as downwards/upwards facing arrows, respectively. The different colors represent different masses and metallicities. The horizontal dotted line shows [X/Fe] = 0.00 dex. For all models, we assume [$\alpha$/Fe] = 0.50 dex and set [Y/Fe] = 0.00 dex.}
	\label{fig:fruity-models}
\end{figure*}

We use the F.R.U.I.T.Y. database \citep{cristallo2011evolution} to compare our abundance measurements with synthetic models of AGB stars. For our models, we assume an initial rotational velocity of zero and adopt the standard configuration for \textsuperscript{13}C pockets. We systematically vary the mass of the AGB stars while fixing the metallicity at Z = 0.00005 and set [$\alpha$/Fe] = 0.50 dex. This metallicity corresponds to [Fe/H] $\simeq$ $-$2.40 dex, which is the closest available value in the F.R.U.I.T.Y. database to our computed metallicity in Sect. \ref{section:stellar-parameters}: [Fe/H] $\simeq$ $-$2.60 dex, Z = 0.0000682093. The conversion follows the relation Z = Z\textsubscript{solar} $\cdot$ 10\textsuperscript{[Z/H]}, where we adopt a solar metallicity of Z\textsubscript{solar} = 0.013 based on \citet{grevesse2012new}. Additionally, we generate models for the same mass range at an even lower metallicity of  Z = 0.00002, [$\alpha$/Fe] = 0.50 dex, corresponding to [Fe/H] $\simeq$ $-$2.81 dex. We present the resulting plot in Fig. \ref{fig:fruity-models}.

Our analysis suggests that the models do not match any current nucleosynthesis model, indicating that AGB stars cannot explain the observed abundance pattern. The lower abundance of Ba suggests a lack of significant AGB star enrichment, consistent with expectations for low-metallicity stars or environments with minimal s-process contribution, which we can confirm when comparing to the model.

By comparing all observed abundances to the F.R.U.I.T.Y models, we can conclude that additional stellar sources or nucleosynthetic processes may have played a role in shaping the chemical composition of this star. Potential contributors include e.g., rotating massive stars \citep{frischknecht2012non}, i-process nucleosynthesis, $\nu$p-process, NSMs, or CC SNe, each of which could have played a role in shaping the chemical enrichment of this star.

\end{appendix}

\end{document}